\documentclass[aps,prl,twocolumn,superscriptaddress]{revtex4}

\usepackage{color}
\usepackage{graphicx}

\begin{document}


\title{Vortex avalanches in the non-centrosymmetric superconductor Li$_2$Pt$_3$B }



\author{C. F. Miclea}
\email[]{miclea@cpfs.mpg.de}
\affiliation{Max-Planck-Institute for Chemical Physics of Solids, Dresden, Germany}

\author{A. C. Mota}
\affiliation{Max-Planck-Institute for Chemical Physics of Solids, Dresden, Germany}
\affiliation{Solid State Lab., ETH-Zurich, Zurich, Switzerland}

\author{M. Sigrist}
\affiliation{Institute for Theoretical Physics, ETH-Zurich, Zurich, Switzerland}

\author{F. Steglich}
\affiliation{Max-Planck-Institute for Chemical Physics of Solids, Dresden, Germany}

\author{T. A. Sayles}
\affiliation{Department of Physics and Institute for Pure and Applied Physical Sciences,
University of California, San Diego, La Jolla, California, USA}

\author{B. J. Taylor}
\affiliation{Department of Physics and Institute for Pure and Applied Physical Sciences,
University of California, San Diego, La Jolla, California, USA}

\author{C. McElroy}
\affiliation{Department of Physics and Institute for Pure and Applied Physical Sciences,
University of California, San Diego, La Jolla, California, USA}

\author{M. B. Maple}
\affiliation{Department of Physics and Institute for Pure and Applied Physical Sciences,
University of California, San Diego, La Jolla, California, USA}


\date{\today}

\begin{abstract}
We investigated the vortex dynamics in the non-centrosymmetric superconductor
Li$_2$Pt$_3$B in the temperature range 0.1~K~--~2.8~K. Two different logarithmic creep
regimes in the decay of the remanent magnetization from the Bean critical state have been
observed. In the first regime, the creep rate is extraordinarily small, indicating the
existence of a new, very effective pinning mechanism.  At a certain time a vortex
avalanche occurs that increases the logarithmic creep rate by a factor of about 5 to 10
depending on the temperature. This may indicate that certain barriers against flux motion
are present and they can be opened under increased pressure exerted by the vortices. A
possible mechanism based on the barrier effect of twin boundaries is briefly discussed.
\end{abstract}


\maketitle



The occurrence of superconductivity in compounds with non-centrosymmetric crystal
structures has attracted considerable attention recently. Besides various other systems,
superconductivity has also been reported in the ternary boride compounds Li$_2$Pd$_3$B
and Li$_2$Pt$_3$B which have superconducting critical temperatures of 7-8 K and 2.4 K,
respectively \cite{Togano,Badica}. These two isostructural compounds crystallize in a
structure consisting of distorted boron centered octahedra of BPd$_6$ or BPt$_6$ in an
approximately cubic arrangement with an inter-penetrating lithium formation
\cite{Eibenstein}. Both substructures, and hence the composite crystal structure, lack
inversion symmetry. Several unusual properties appear in non-centrosymmetric
superconductors depending upon various factors, in particular the specific form of the
spin-orbit coupling in such systems as well as the pairing symmetry
\cite{Gorkov,Frigeri,Edelstein}. In contrast to the strongly correlated
non-centrosymmetric heavy fermion superconductors CePt$_3$Si \cite{Bauer}, CeRhSi$_3$
\cite{Kimura}, and UIr \cite{Akazawa} for which superconductivity is associated with a
magnetic quantum phase transition, there is no evidence of magnetic order or strong
electron correlations in either Li$_2$Pd$_3$B or Li$_2$Pt$_3$B. Measurements of the
London penetration depth suggest that Li$_2$Pd$_3$B has a full quasiparticle gap in the
superconducting phase, while for Li$_2$Pt$_3$B, the data indicate line nodes in the
energy gap  \cite{Yuan}. NMR measurements \cite{Nishiyama} suggest that Li$_2$Pd$_3$B is
a spin singlet, s-wave superconductor. In contrast, in Li$_2$Pt$_3$B, the spin
susceptibility measured by the Knight shift remains unchanged across the superconducting
transition temperature, and the spin-lattice relaxation rate $1/T_1$ shows no coherence
peak below $T_c$, decreasing as $T^3$ with decreasing temperature, consistent with gap
line nodes. In this letter, we investigate a further intriguing property of the
unconventional superconductor Li$_2$Pt$_3$B, observed in the vortex dynamics. We
demonstrate that the behavior of the flux creep is very unusual, displaying at short
times extremely small creep rates followed by a faster avalanche-like escape of magnetic
flux.

Samples of Li$_2$Pt$_3$B used in this experiment were synthesized in an arc furnace
utilizing a two-step process similar to that outlined in the work of Badica et al.
\cite{Badica}. An initial binary sample of BPt$_3$ was grown using Pt of purity 99.99 \%
and B of purity 99.999 \%. In the final step of sample growth, an excess amount of Li was
added in order to account for losses during arc melting, giving a Li to BPt$_3$ ratio of
2.2:1. The crystal structure was verified via powder X-ray diffraction measurements. No
impurity or binary phases were detected.

Prior to the magnetic relaxation measurements, the Li$_2$Pt$_3$B sample  was
characterized by means of measurements of electrical resistivity $\rho$, magnetization
$M$, and specific heat $C$. All three measurements yielded a value of the superconducting
critical temperature $T_c = 3.0$~K. This value is significantly higher than the values
reported in the literature, the highest of which is $T_c = 2.4$~K \cite{Badica}.
Evidently, the value of $T_c$ is rather sensitive to the composition of the sample, which
could be a further indication that Cooper pairing is unconventional in this compound.

The temperature dependence of the specific heat was measured using a quasi-adiabatic heat
pulse method in a $^3$He refrigerator, in the temperature range of $0.6~$K$\leq T \leq
20$~K, and in magnetic fields up to $H \approx 4$~T.  The temperature dependence of the
specific heat in the normal state could be described by the expression $C(T) = C_e(T) +
C_l(T)$, where $C_e(T)=\gamma T$ is the electronic contribution and $C_l(T)=\beta T^3$ is
the lattice term.  The best fit to the data yields $\gamma = 7.0$~mJ/(mol~K$^2$) for the
electronic specific heat coefficient and $\Theta_D = 203$~K for the Debye temperature, in
good agreement with the values $\gamma = 7.0$~mJ/(mol~K$^2$) and $\Theta_D = 228$~K
reported by Takeya et al. \cite{Takeya}. A relatively sharp jump in the specific heat,
$\Delta C=15$~mJ/(mol K), was observed at the transition into the superconducting state,
yielding a ratio  $\Delta C/\gamma T_c = 1.2$, larger than the value of 0.8 reported by
Takeya et al. \cite{Takeya} but smaller than the weak coupling BCS value of 1.43. Below
$T_c$, $C_e(T)$ decreases nearly as $T^2$ upon decreasing the temperature to $T \approx
0.5~T_c$, consistent with the behavior reported previously \cite{Takeya}  and in contrast
to the exponential $T$-dependence expected for a BCS superconductor. The upper critical
field, $H_{c2}$, determined from the $C(T)$ measurements, increases linearly with
decreasing temperature from $T_c$ to $T = 0.6$~K and extrapolates linearly to a value of
$H_{c2}(0)\approx 1.5$~T at $T = 0$~K.

\begin{figure}
\includegraphics [height=2.3in] {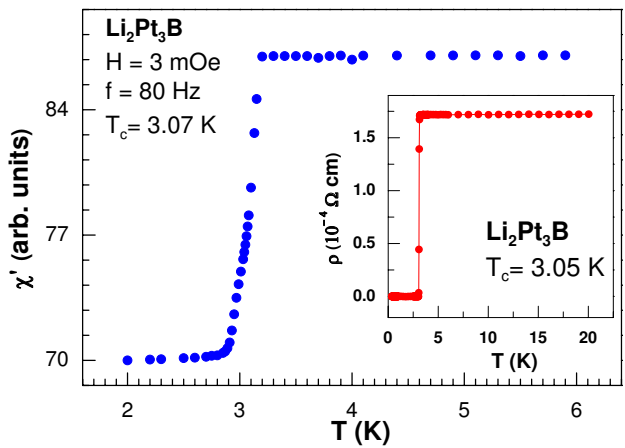}
\caption{\label{fig1} {Real part of the ac magnetic susceptibility, $\chi'$ as a function
of temperature across the superconducting phase transition. Inset: electrical
resistivity, $\rho$, as function of temperature.}}
\end{figure}

We also characterized the superconducting transition of the sample by means of ac
magnetic susceptibility measurements in a low ac field of $H=3$~mOe and at a frequency of
$f=80$~Hz. This measurement was done with the sample situated inside a custom built
mixing chamber of a dilution refrigerator using an inductance bridge with a SQUID as a
null detector \cite{Amman}. The midpoint of the superconducting  transition of this
sample is at $T_c= 3.07$~ K and the transition width   $\Delta T_c =  240$~mK. The data
are displayed in Fig.1. In the inset of Fig. 1, we show the low temperature part of the
electrical resistivity, $\rho (T)$ measured using a standard four wires arrangement in a
$^3$He cryostat. $\rho (T)$ which displays a sharp phase transition into the
superconducting state with a width $\Delta T_c = 55$~mK. $\rho (T)$ reaches zero at $T=
3.05$~K, in very good agreement with the susceptibility data. In the normal state, the
$\rho (T)$ measurement revealed typical metallic behavior.

\begin{figure}
\includegraphics [height=2.3in] {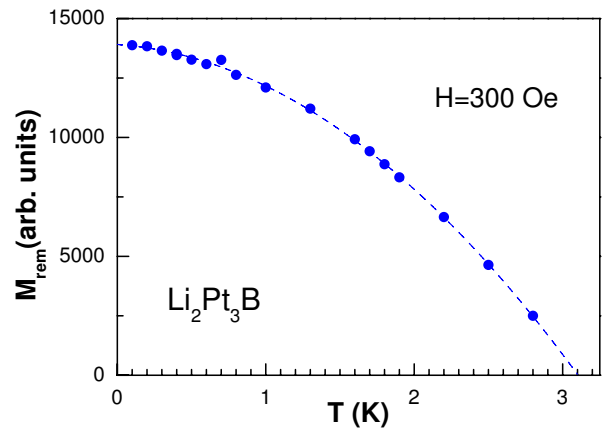}
\caption{\label{fig3} {Temperature dependence of the remanent magnetization
$M_{rem}$}. The dotted line is a parabolic fit to the data.}
\end{figure}

\begin{figure}
\includegraphics [height=2.3in] {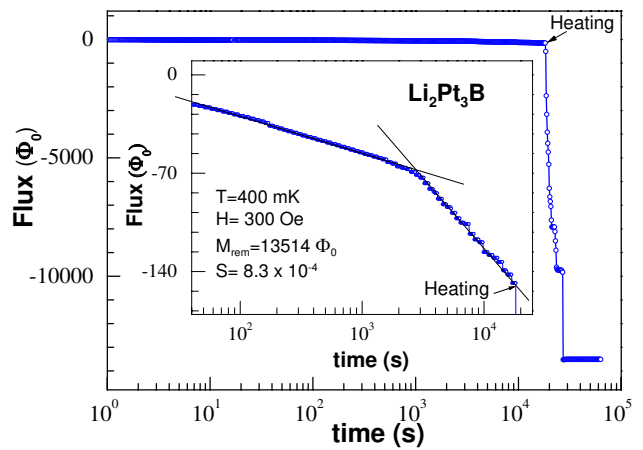}
\caption{\label{fig3} {A typical relaxation curve exemplified for $T=400$~mK. The
magnetic flux is measured as it is expelled out of the sample at constant temperature.
After a certain time (marked by arrows in the figure), the sample is gradually heated and
driven into the normal state. Inset: Two relaxation regimes are visible in the
expanded scale.}}
\end{figure}

The investigation of vortex dynamics was performed in the temperature range 0.1K - 2.8 K.
Isothermal relaxation curves of the remanent magnetization $M_{rem}$ were taken after
cycling the specimen in an external field H. Vortices were introduced into the sample at a
slow rate in order to avoid eddy current heating. After waiting for several  minutes, the
magnetic field was reduced to zero and the relaxation of the metastable magnetization
recorded with a digital flux counter for several hours. The sample was prepared in the
form of a thin slice and the magnetic field was applied along the longest direction. At
$T = 100$~mK, we determined the field corresponding to the Bean critical state for this
sample to be $H=300$~Oe. In Fig. 2, we show values of the remanent magnetization obtained
after cycling the sample in a field of $H=300$~Oe as a function of temperature. $M_{rem}$
decreases monotonically upon increasing the temperature with the experimental data well
fitted by a parabola (dashed line in Fig. 2) which reaches zero at around $T \approx
3.1$~K. This is in excellent agrement with the value of $T_c$ yielded by specific heat,
ac susceptibility and electrical resistivity measurements.
\\

\begin{figure}[!b]
\centering\renewcommand{\baselinestretch}{1}
\includegraphics [width=0.5\textwidth] {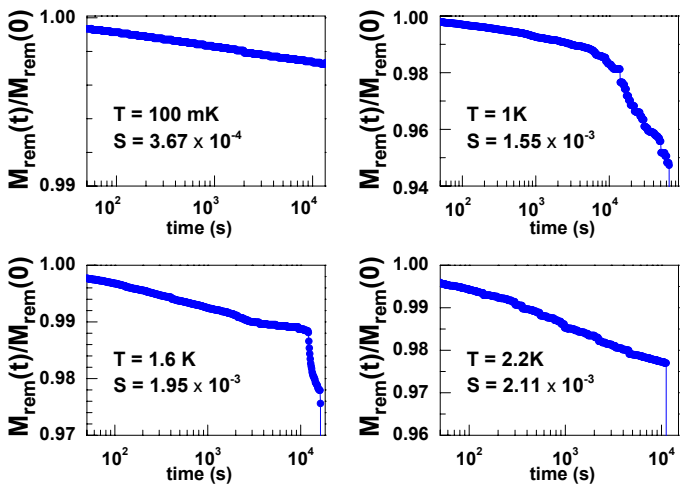}
\caption{\label{fig4} {Sequence of isothermal relaxation curves. Two relaxations
regimes can be observed at intermediate temperatures as explained in the text.} }
\end{figure}

A typical decay of the remanent magnetization from the critical Bean state at $T=400$~mK
is shown in Fig. 3.  In this case, the creep was recorded for about 18000~s. At that
time, the sample was heated above $T_c$ in order to obtain the total value of the
remanent magnetization as a sum of the amount decayed in the first 18000~s plus the
quantity expelled on crossing $T_c$. This value is then used to normalize the creep rate.
In the inset, the same data are displayed on an expanded scale. We can clearly
distinguish two different logarithmic creep regimes. For $50~$s $ < t < 2400$~s we
observe a clear logarithmic relaxation law, with an extremely low relaxation rate,
$S=\partial \ln M/\partial \ln t = 8.3 \times 10^{-4}$. At around $t=2400$~s, a sudden,
strong increase of the relaxation rate occurs, following also a logarithmic law, but with
a rate about a factor of four larger, $S =3.1 \times 10^{-3}$. Indeed, an avalanche-like
escape of vortices  has suddenly occurred around $t=2400$~s, indicating that the
relaxation process is rather complex. Vortices escaping the sample apparently need a
considerable amount of time to overcome a certain barrier. We observed this type of
regime change at all temperatures (see Fig. 4), except for relaxations below $T=400$~mK
and above $T=2$~K.
\begin{figure}
\includegraphics [width=0.45\textwidth] {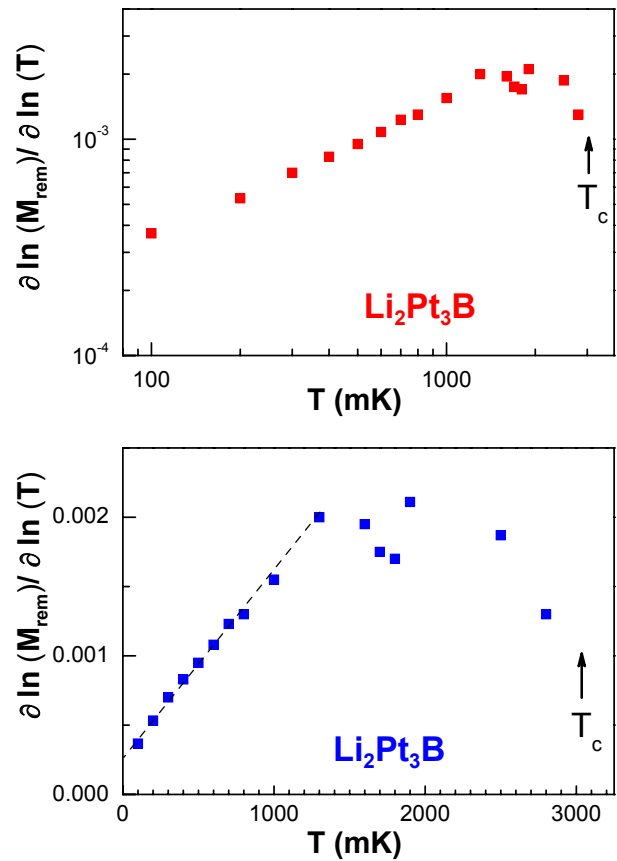}
\caption{\label{fig5} {The temperature dependence of the initial decay rate $S=\partial \ln
M / \partial \ln t$ in a double logarithmic (upper panel) and a linear plot (lower
panel).}}
\end{figure}
To our knowledge, this phenomenon of avalanches in the slow decay of vortices toward
equilibrium has never been observed before in any superconductor, conventional or
unconventional. This unexpected result points to new vortex physics in this
non-centrosymmetric superconductor. The normalized relaxation rates corresponding to the
creep before the avalanches occur are depicted in Fig. 5. In the upper panel, the rates
are given in a double logarithmic plot while, on the lower one, in linear scales. These
rates are even lower by a factor of five than the very weak creep rates observed in
PrOs$_4$Sb$_12$  \cite{Cichorek}, a superconductor that violates time reversal symmetry.
As discussed by Sigrist and Agterberg  \cite{Sigrist}, the lack of time reversal symmetry
in such superconductors allows for the formation of flux flow barriers formed by
fractional vortices on domain walls of the superconductor, that can prevent the motion of
normal vortices.

 A possible explanation for the extremely slow motion of flux lines in Li$_2$Pt$_3$B
has recently been given by Iniotakis et al \cite{Iniotakis}. In many cases for
non-centrosymmetric materials, the absence of an inversion center allows for the twinning
of the crystal. These authors have shown theoretically, that a phase that violates time
reversal symmetry can also be realized at interfaces separating crystalline twin domains
of opposite spin-orbit coupling. In this case, flux lines with fractional flux quanta
could exist on such interfaces and turn twin boundaries into strong barriers impeding
flux creep. Within this model, vortex avalanches could be expected when such a fence
opens due to excessive pressure of normal vortices. This is possible, if the vortex
density increases to a level such that fractional vortices can no longer exist and the
vortex pinning mechanism of twin boundaries fails. In the frame of this scenario, we can
interpret the temperature window in which the avalanche effect has been observed in the
following way. At temperatures below $T=400$~mK, vortices move so slowly that the time
necessary to build up the necessary vortex density to break a barrier exceeds the
observation time. For temperatures above $T=2$~K, on the other hand, the overall density
of vortices is strongly reduced
(see Fig. 2), so that it becomes more difficult to reach the density required for demolishing  barriers.

In conclusion, we have observed in Li$_2$Pt$_3$B strong avalanches in the relaxation of
the remanent magnetization. Prior to the avalanches, vortices move toward equilibrium for
several hours with an extraordinary slow creep rate, indicating that a new type of
pinning is effective in this non-centrosymmetric superconductor. This type of pinning is
different from the conventional pinning by defects, since it is effective only at low
density of vortices. If the density of vortices leaving the sample increases close to the
barrier keeping them from moving, an avalanche occurs followed by creep rates that are
about 5 to 10 times faster. This creep behavior has never been observed up to now. A
possible mechanism has been proposed by Iniotakis et al.,  based on the role of twin
boundaries in non-centrosymmetric superconductors \cite{Iniotakis}.

We would like to thank to C. Iniotakis and S. Fujimoto for helpful discussions. C.F.M
would like to acknowledge the support of the German Research Foundation (DFG) under the
auspices of the MI 1171/1-1. A-C.M. and M.S. are grateful for financial support from the
Swiss Nationalfonds and the NCCR MaNEP. The work done at the University of California was
supported by the US Department of Energy under grant no. DE FG02-04ER46105.

\end{document}